\documentclass[sigconf]{acmart}
\AtBeginDocument{%
  \providecommand\BibTeX{{%
    \normalfont B\kern-0.5em{\scshape i\kern-0.25em b}\kern-0.8em\TeX}}}

\setcopyright{acmcopyright}

\copyrightyear{2024}
\acmYear{2024}
\setcopyright{rightsretained}
\acmConference[FPGA '24]{Proceedings of the 2024 ACM/SIGDA International Symposium on Field Programmable Gate Arrays}{March 3--5, 2024}{Monterey, CA, USA}
\acmBooktitle{Proceedings of the 2024 ACM/SIGDA International Symposium on Field Programmable Gate Arrays (FPGA '24), March 3--5, 2024, Monterey, CA, USA}
\acmDOI{10.1145/3626202.3637558}
\acmISBN{979-8-4007-0418-5/24/03}

\usepackage{siunitx}

\usepackage{xspace}

\newcommand{\aname}{ESDA\xspace}
\newcommand{\figref}[1]{\figurename~{\ref{#1}}}
\newcommand{\tabref}[1]{Table~{\ref{#1}}}
\newcommand{\secref}[1]{Section~{\ref{#1}}}
\newcommand{\eqnref}[1]{Eqn.~{\ref{#1}}}

\usepackage{multirow}

\begin{document}

\title{A Composable Dynamic Sparse Dataflow Architecture for Efficient Event-based Vision Processing on FPGA}

\author{Yizhao Gao}
\email{yzgao@eee.hku.hk}
\orcid{0000-0001-5673-3746}
\affiliation{%
  \institution{University of Hong Kong}
  \country{Hong Kong}
}

\author{Baoheng Zhang}
\email{bhzhang@eee.hku.hk}
\orcid{0000-0003-1685-656X}
\affiliation{%
  \institution{University of Hong Kong}
  \country{Hong Kong}
}

\author{Yuhao Ding}
\email{yhding@eee.hku.hk}
\orcid{0009-0009-4819-9251}
\affiliation{%
  \institution{University of Hong Kong}
  \country{Hong Kong}
}

\author{Hayden Kwok-Hay So}
\email{hso@eee.hku.hk}
\orcid{0000-0002-6514-0237}
\affiliation{%
  \institution{University of Hong Kong}
  \country{Hong Kong}
}

\begin{abstract}
 Event-based vision represents a paradigm shift in how vision information is captured and processed. By only responding to dynamic intensity changes in the scene, event-based sensing produces far less data than conventional frame-based cameras, promising to springboard a new generation of high-speed, low-power machines for edge intelligence.
However, processing such dynamically sparse input originated from event cameras efficiently in real time, particularly with complex deep neural networks (DNN), remains a formidable challenge. Existing solutions that employ GPUs and other frame-based DNN accelerators often struggle to efficiently process the dynamically sparse event data, missing the opportunities to improve processing efficiency with sparse data.
To address this, we propose \aname, a composable dynamic sparse dataflow architecture that allows customized DNN accelerators to be constructed rapidly on FPGAs for event-based vision tasks.  \aname is a modular system that is composed of a set of parametrizable modules for each network layer type.  These modules share a uniform sparse token-feature interface and can be connected easily to compose an all-on-chip dataflow accelerator on FPGA for each network model.
To fully exploit the intrinsic sparsity in event data, \aname incorporates the use of submanifold sparse convolutions that largely enhance the activation sparsity throughout the layers while simplifying hardware implementation.
Finally, a network architecture and hardware implementation co-optimizing framework that allows tradeoffs between accuracy and performance is also presented.
Experimental results demonstrate that when compared with existing GPU and hardware-accelerated solutions, \aname achieves substantial speedup and improvement in energy efficiency across different applications, and it allows much wider design space for real-world deployments.

\end{abstract}







\maketitle

\section{Introduction}

Event-based vision systems are bio-inspired systems that represent a paradigm shift from conventional frame-based vision systems~\cite{event-based-vision}.
Instead of capturing the entire field of view as dense matrices of pixel values like in a conventional camera, an event camera only produces a spiking event when a substantial change in light intensity local to each pixel is detected.
Despite the fundamentally different way of sensing, substantial progress has been made in recent years to perform a wide range of vision tasks such as object classification, object detection, and tracking utilizing deep neural network models~\cite{deep_event_survey}. 

However, deploying event-based vision systems with deep learning models in near-sensor or edge-computing contexts remains a great challenge. 
On the one hand, off-the-shelf GPU platforms and dense DNN accelerators that are optimized to operate on dense input from conventional cameras can rarely exploit the sparsity in event-based inputs.
As a result, if blindly employed to accelerate event camera output, these systems struggle to deliver the promised energy efficiency and performance benefits due to a large amount of wasted computation.
On the other hand, some sparse DNN accelerators have previously been proposed to address the need for event vision processing~\cite{nullhop, nullhop_fpga}.
Yet, these designs primarily concentrate on optimizing sparse computation at the matrix multiplication or single layer level, often through a reusable processing engine for various layers. Such an approach may lead to considerable latency and energy overhead, mainly due to the recurrent input/output operations involving weights and intermediate activations.
Consequently,  these designs are suboptimal for real-time fast-reacting systems intended to exploit the low latency sensing and high temporal resolution benefits of event cameras.

To address these challenges, we introduce \aname, a system based on FPGA that can efficiently accelerate \textbf{E}vent-based vision processing with a composable dynamic \textbf{S}parse \textbf{D}ataflow \textbf{A}rchitecture.
\aname is a modular system that allows parametrizable modules for each network layer to be connected together to form a customized all-on-chip system on FPGA.
Employing the all-on-chip dataflow architecture is a key design choice towards low latency and power consumption by minimizing the off-chip communication of weight and activation.
In contrast to prior dataflow architectures that handle dense frame-based input, \aname exclusively streams the spatially sparse features from the event-based input across the different dataflow modules by using a unified sparse token-feature interface. 
Through the straightforward cascading of dataflow modules corresponding with the model definition, a sparse accelerator can be rapidly constructed to speed up the application. 
Furthermore, to take full advantage of the sparse dataflow approach, \aname also integrates submanifold sparse convolution~\cite{submanifold} at the algorithmic level. Submanifold convolution can effectively maintain high spatial sparsity of feature activations of the models by constraining the spatial activation locations, which also helps to derive simpler hardware implementation.  

\begin{figure*}[t]
    \includegraphics[width=1\linewidth]{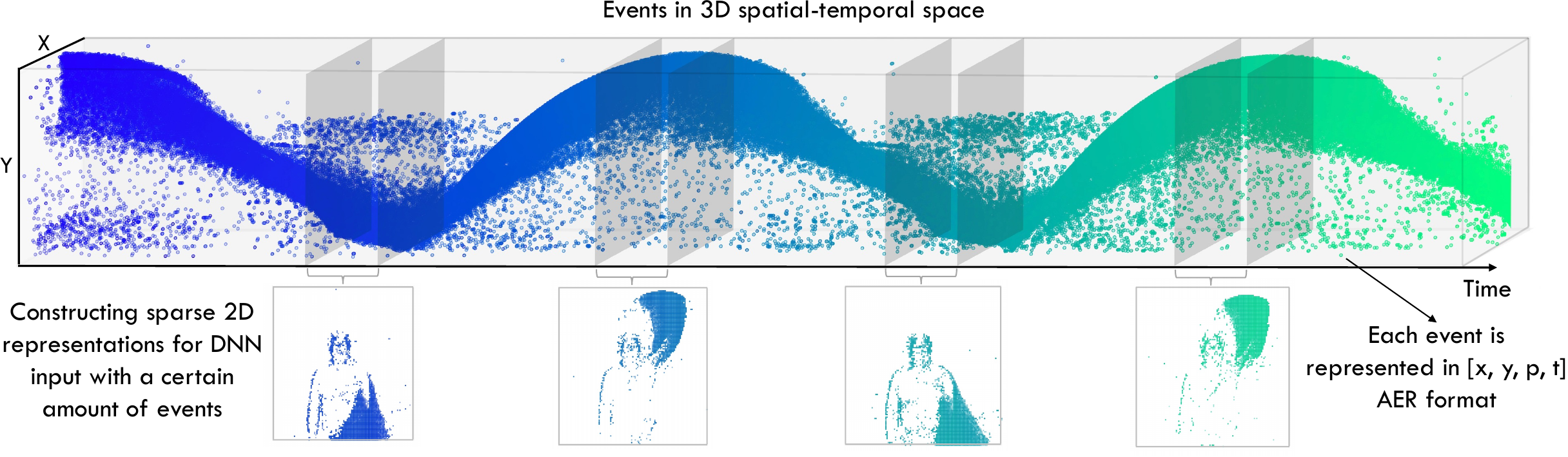}
    \centering
    \caption{Working principle of event camera. The event camera only captures the light intensity change as spiking events in AER format ([x, y, ±1, timestamp]). This figure shows a recording sample from DvsGesture dataset where a man is rotating his left arm counter-clockwise. Due to the dynamic nature of event cameras, only the motions of the man are captured. For vision tasks like object recognition, a certain amount of events are usually grouped to form a 2D representation as DNN input.}     
    \label{fig:event_principle}
\end{figure*}

To facilitate design space exploration, we also develop a network model and hardware architecture co-optimization framework that leverages the sparsity statistics from a given event-based vision task or dataset to perform accuracy and hardware efficiency tradeoffs. Given the capacity of today's FPGA edge device, the search framework can identify many efficient models that fit on-chip while achieving good performance and accuracy.
To this end, the main contribution of this work can be summarized as:
\begin{itemize}
\item We proposed a novel composable sparse dataflow architecture for DNN inference in event-based vision applications that effectively harness the inherent sparsity in event camera data. 
\item We implemented \aname on embedded FPGA platforms and established a set of sparsity-aware co-optimization flow, enabling the customization of DNN accelerators for each specific application or dataset.
\item We showcased a substantial end-to-end hardware speedup and energy-efficiency improvement through extensive evaluation on multiple event-based datasets.
\end{itemize}

In the next section, background and related work about neuromorphic vision systems and processing will first be provided. The detailed design of \aname will be discussed in \secref{sec:methodology}. Experiment results are shown in \secref{sec:evaluations} and we will conclude in \secref{sec:conclusion}.  The artifact associated with this work is open-sourced and available at~\cite{ecda_artifact}.

\section{Background and Related Work}

\subsection{Event-based Vision}
Event cameras, also known as Neuromorphic Cameras or Dynamic Vision Sensors, are designed to mimic the functioning of the human retina. Instead of capturing full frames, they detect and report light intensity changes as events asynchronously with high temporal resolution. Events are usually encoded in Address Event Representation (AER) format. Typically each event can be written as $[x,y,p,t]$ where $x,y$ points to the location of the event, $t$ is the timestamp and $p$ is the polarity of changes. Commercial event cameras nowadays can support $720\times 1280$ spatial resolution with \SI{1}{\micro\second} temporal resolution~\cite{prophesse_camera}. 
Compared with traditional frame-based cameras, event cameras have several advantages, including: (1) high temporal resolution allowing capturing fast events that conventional cameras may miss; (2) high dynamic range enabling them to capture scenes with varying lighting conditions; (3) high efficiency in terms of data processing and transmission. 
As an emerging technology, event-based vision has shown great potential in areas like robotics, autonomous vehicles, high-speed tracking, scientific imaging, low-power embedded systems, etc~\cite{event-based-vision}.    

\figref{fig:event_principle} shows a sample from event-based dataset DvsGesture~\cite{dvsgesture} to illustrate the working principle of the event camera. 
In this sample, a man is rotating his arm counter-clockwise. 
As the event camera exclusively responds to changes in scene intensity, it captures only the movements of the person as a sequence of events. The upper figure shows the scatter plots of events in 3D spatial-temporal space. The below four images show the 2D feature representations constructed from events in a short time interval. It's a common practice to use these 2D representations to serve as the DNN input.
Prior research works have introduced many different 2D representation algorithms like event histogram~\cite{histogram_and_autodrive}, time-surface~\cite{hots, hats}, learnable MLP kernels~\cite{mlp_repre}, and etc.
It's worth noting that \aname can seamlessly integrate with different 2D representation algorithms. As long as the resulting feature is spatially sparse, it can benefit from the sparse architecture of \aname.



\subsection{Hardware Acceleration of Event-based Vision}

From a system perspective, fully unlocking the potential of the event-based approaches cannot be achieved without considering efficient hardware processing.
Previous efforts have mainly been devoted to two directions.  
On the one hand, due to the similar neuromorphic working principle, many research works have integrated event cameras with bio-inspired neuromorphic computing devices. Typically, these works seek to accelerate Spiking Neural Networks (SNNs) on customized hardware such as TruthNorth~\cite{truenorth_chip}, Loihi~\cite{loihi}, SpiNNaker~\cite{spinnaker}, and etc. However, training SNN for some complex event-based vision tasks itself can also be challenging. Moreover, despite deployment on neuromorphic hardware like TrueNorth, the latency for network inference remains substantial, e.g. \SI{105}{ms} on DVS Gesture Recognition.  

On the other hand, some domain-specific architectures for event-based vision tasks have been proposed, such as optical flow calculation~\cite{opticalflow0, opticalflow1, opticalflow12}, object tracking~\cite{remot, remot_trans, tracking_0, hardware_tracking_1}, feature recognition~\cite{feature, corner_det}, etc.
In recent years, the event-based vision has seen significant advancements through integrating deep neural networks (DNNs) in its processing pipeline~\cite{deep_event_survey}. However, accelerators originally designed for frame-based DNN models struggle to efficiently process event-based inputs with the required efficiency.
Many sparse DNN accelerators have primarily emphasized element-wise weight sparsity or activation sparsity induced by the ReLU~\cite{scnn, eyerissv2, Cambricon-S, SparTen}. These approaches differ significantly from the intrinsic spatial sparsity in event data.
While some accelerators can leverage spatial sparsity, they often target at point cloud~\cite{pointacc, crescent} or dynamic spatial pruning networks~\cite{dpcas, spatial_pruning_hardware}, which are also different in terms of the source and pattern of sparsity.

Thus, there is a growing need for specialized accelerators to effectively leverage the unique spatial sparsity characteristics of event-based data to unlock its advantages in real-time energy-efficient DNN inference. Prior work like NullHop~\cite{nullhop, nullhop_fpga} has proposed an architecture that uses a binary bitmap to represent layer-wise sparse activation to skip the computation of zeros. Other solutions like PFF~\cite{ppfbnn} have also explored treating the event-based input sparsity as a bit-level sparsity by using Binary Neural Networks. 
In this work, we aim to achieve low-latency and energy-efficient processing of event-based DNN models by proposing an all-on-chip sparse dataflow architecture on FPGA. The framework is also highly composable, allowing the customization of accelerators for different event-based models and datasets.

\section{Methodology} \label{sec:methodology}

\subsection{Design Principle of Sparse Dataflow}

\aname is built upon distinct sparse dataflow modules that share a unified token-feature interface. The framework enables seamless customization of a sparse accelerator by cascading various modules aligned with the model's architecture. By definition, \aname only streams spatially sparse features along with their coordinate tokens through the dataflow modules. 
\eqnref{eq:dataflow} shows the design principle for an \aname dataflow module using a unified sparse token-feature interface:
\begin{equation}
\begin{split}
    \text{token}_\text{{out}}[i] &= f_T(\text{token}_\text{{in}}[j],...,\text{token}_\text{{in}}[j-N]) \\
    \text{feature}_\text{{out}}[i] &= f_F(\text{feature}_\text{{in}}[j],...,\text{feature}_\text{{in}}[j-N]) \\
    \text{s.t.} \ \ ravel(&\text{token}_\text{{out}}[i]) > ravel(\text{token}_\text{{out}}[i - 1])
\end{split}
\label{eq:dataflow}
\end{equation}
where each $token[i]$ (written in $[.x, .y, .end]$) marks the spatial location of the current feature vector $feature[i]$. The flag $.end$ denotes whether it's the end of the stream. The $ravel$ order, calculated as $(y * Width + x)$, refers to the left-to-right, top-to-bottom streaming order of spatial locations, which also corresponds to the 1D memory layout order of a 2D dense matrix. In other words, the module itself should maintain the correct output order so that different modules can be pipelined. 

The functionality of each dataflow module is separated into: (1) $f_T$ to resolve the new output token based on previous input token information; (2) $f_F$ to compute the output features at new output token locations by using some temporal neighboring features. Depending on the algorithm requirement, different numbers of temporal neighboring data might be needed, specified as $N$ in \eqnref{eq:dataflow}. Each module in \aname is designed as a flexible hardware template, where many hardware design parameters like bitwidth, parallel factor, buffer size, etc., can be specified at compile time.

\begin{figure}[t]
    \includegraphics[width=\linewidth]{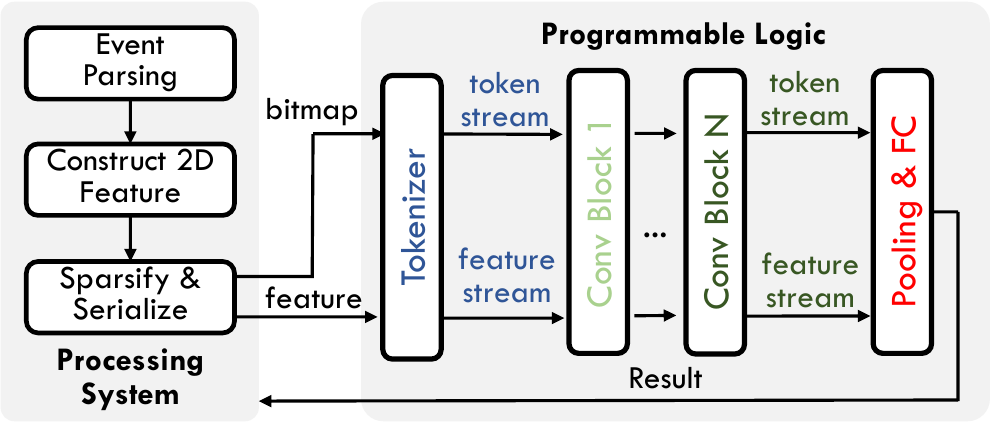}
    \centering
    \caption{Overall architecture of an ESDA accelerator.}
    \label{fig:hardware diagram}
\end{figure}

\figref{fig:hardware diagram} shows an illustrative accelerator diagram for the event-based object classification task. The processing system, or CPU, is responsible for event streaming and 2D representation construction. While the accelerator on the programmable logic reads the sparse input features following the left-to-right, top-to-bottom ordering to align with our dataflow requirement. A binary bitmap is used to mark the non-zero locations and generate the token stream. The tokenized sparse feature then propagates through each convolution block. Finally, the accelerator returns the classification outputs back to the processing system.


In the following section, we will first discuss how \aname incorporates submanifold sparse convolution, which is used to enhance the model's sparsity while simplifying the dataflow design ($f_T$, $f_F$ and $N$ in \eqnref{eq:dataflow}). In \secref{sec:sparse architecture}, we will show the detailed micro-architecture of some commonly used dataflow modules.


\subsection{Sparsify the Algorithm: Submanifold Sparse Convolution}
\label{sec:sparse algorithm}
\begin{figure}[t]
    \includegraphics[width=1\linewidth]{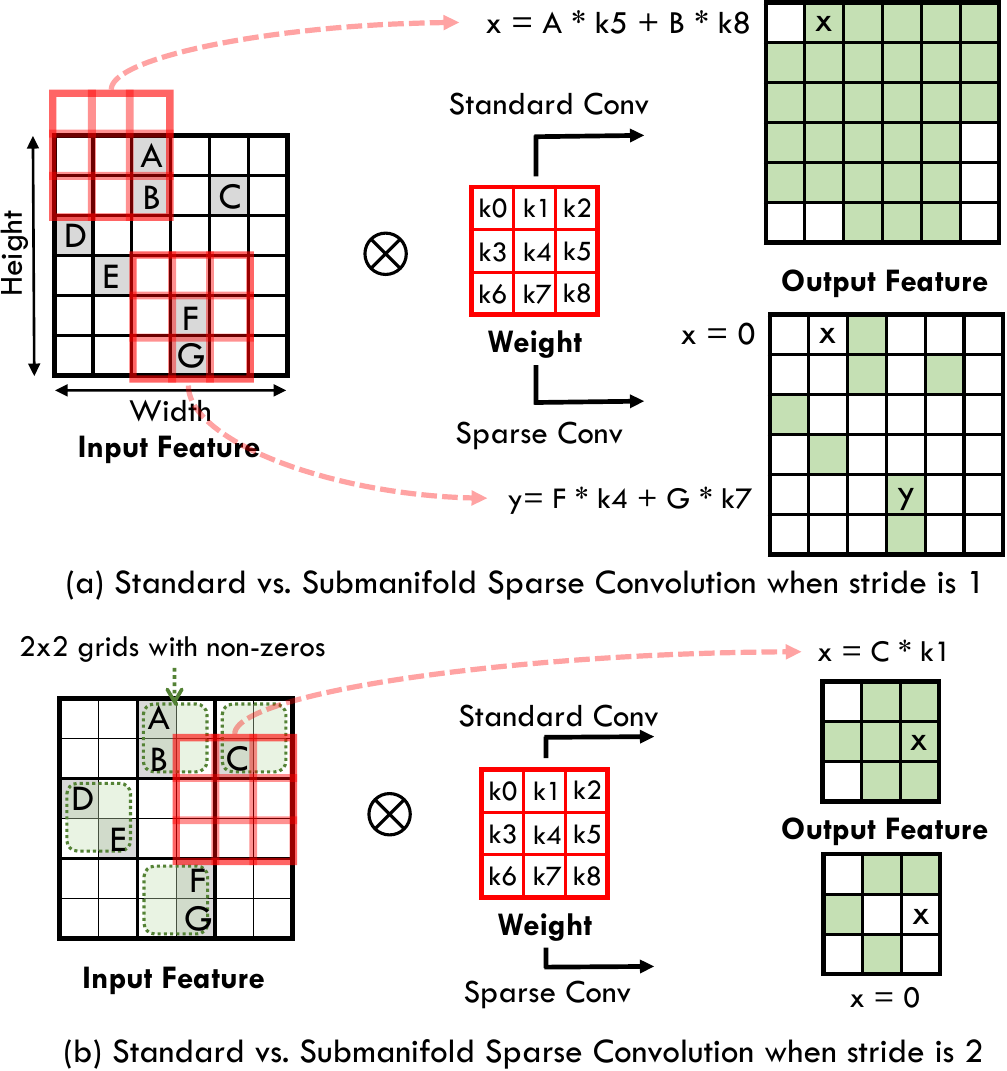}
    \centering
    \caption{Compare standard convolution with submanifold sparse convolution. The gray/green locations in the figure mean non-zero pixels. (a) When stride $s = 1$, the input and output location of submanifold convolution is restricted to be identical. In the "x" location, standard convolution leads to a valid non-zero output while submanifold convolution does not. (b) When stride $s>1 $ (2 in the figure), an output location is non-zero if the corresponding input $s\times s$ grid contains non-zeros.  
    }
    \label{fig:submanifold}
\end{figure}

The input of a DNN model for event-based vision tasks is typically a spatially sparse 2D feature/representation generated by a certain amount of events.  
Although input features can be highly sparse (e.g. <\%1 in ASL-DVS dataset), using standard convolution can easily leads to much denser intermediate features because of the “dilation" effect of convolution. \figref{fig:submanifold} shows the example of sparse input features convoluted with a $3\times 3$ kernel.
The output feature of standard convolution is much denser than the input feature, which can diminish the opportunities brought by the sparsity.

To address this, \aname incorporates submanifold sparse convolution~\cite{submanifold} to better enhance the spatial sparsity of a model. \figref{fig:submanifold} illustrates the operation of submanifold sparse convolution. When the convolution stride $s$ is 1, the submanifold sparse convolution enforces input and output non-zero locations to be identical. This means that for positions that are originally zeros, e.g. "x" in the \figref{fig:submanifold}.a, the output remains zeros. While for those non-zeros output locations, e.g. "y" in the figure, the feature is calculated by using the similar weighted sum algorithm as standard convolution. In this way, submanifold convolution prevents the "dilation" effect of standard convolution by restricting the possible non-zero output locations. 

When convolution stride $s$ is larger than 1, an output location of sparse convolution is non-zero if the corresponding input $s\times s$ stride grid contains any non-zeros. For instance, the "x" in \figref{fig:submanifold}.b remains zero because its corresponding $2\times2$ grid of the inputs are all zeros.
This is in contrast to standard convolution which generates non-zero outputs whenever the convolution kernel meets any non-zeros. Thus, sparser output features can be expected by using sparse convolution if the stride size is smaller than the kernel size, which is a common configuration.  

Submanifold sparse convolution also comes with a simpler hardware implementation. For example, in the stride-1 case where the input-output sparse locations are the same, the input tokens can simply be queued in a FIFO and reused for the output. Consequently, the abstract module function can be written as:
\begin{equation}
\begin{split}
    &\text{token}_\text{{out}}[i] = \text{token}_\text{{in}}[i] \\
    \text{feature}_\text{{out}}[i] &= f_F(\text{feature}_\text{{in}}[i],...,\text{feature}_\text{{in}}[i-N]) 
\end{split}
\end{equation}
In this way, the hardware can co-iterate input/output tokens and features, while saving the additional logic to generate potential output locations during runtime.

\subsection{Sparsify the Architecture: Composable Sparse Dataflow Modules}
\label{sec:sparse architecture}

\subsubsection{$1\times 1$ Convolution Module}
\begin{figure}[h]
    \includegraphics[width=0.83\linewidth]{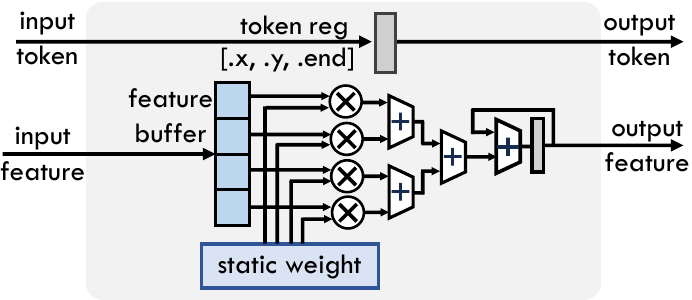}
    \centering
    \caption{Convolution $1\times 1$ Module.}
    \label{fig:conv1x1}
\end{figure}
$1\times 1$ convolution, also known as point-wise convolution, only operates locally by iterating each sparse location and updating its features. Benefiting from this property, the input and output tokens can simply be relayed using a token register.
\figref{fig:conv1x1} shows the hardware diagram of the $1\times 1$ convolution module. The feature vector at the current location $[.x, .y]$ is first read into an input feature buffer and then multiplied with the weight matrix in the static buffer. Since the convolution layers are spatially mapped on the accelerator, the weights will be synthesized into Read-Only Memory using Block RAM (BRAM) and initialized during bitstream download. 
The computation of the output feature vector can be implemented as a Matrix-Vector multiplication unit as illustrated in \figref{fig:conv1x1}. The size of the PE array, which can also be regarded as the parallel factor (PF) on channel dimension, is exposed as a hardware template parameter. These parallel factors can also affect the BRAM usage by partitioning the weights buffer to match the bandwidth requirement for the MAC array. 
Note that the batch normalization layer can be folded into the convolution layer, and the activation function is omitted from the diagram for simplicity.  

\subsubsection{$k\times k$ Convolution Module ($k>1$)}

\begin{figure}[h]
    \includegraphics[page=2, width=0.78\linewidth]{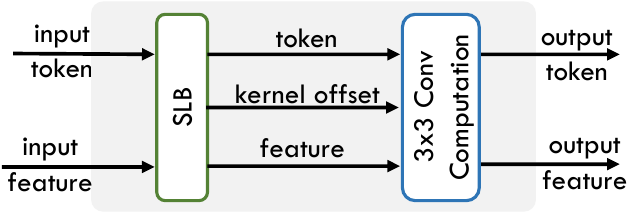}
    \centering
    \caption{Convolution $3\times 3$ Module.}
    \label{fig:conv3x3}
\end{figure}
Different from $1\times1$ convolution that directly operates on non-zero positions locally, when the convolution kernel size $k$ is larger than 1, the computation is the weighted summation of neighboring features. 
In \aname, $k\times k$ convolution is implemented by cascading a Sparse Line Buffer module and a $k\times k$ convolution computation module as shown in \figref{fig:conv3x3}. 
The Sparse Line Buffer designs can be slightly different for different convolution strides (usually 1 or 2), while the computation module remains the same. 
As shown in \figref{fig:submanifold}, the weighted sum of spatial kernels can also be sparse (e.g., "y" only requires $F * k4 + G * k7$). 
To leverage this type of sparsity, an additional kernel offset stream is used to denote the sparse neighboring offset within the $k\times k$ kernel. 

\subsubsection{$k\times k$ Convolution Computation Module ($k>1$)}
\begin{figure}[h]
    \includegraphics[page=3, width=0.92\linewidth]{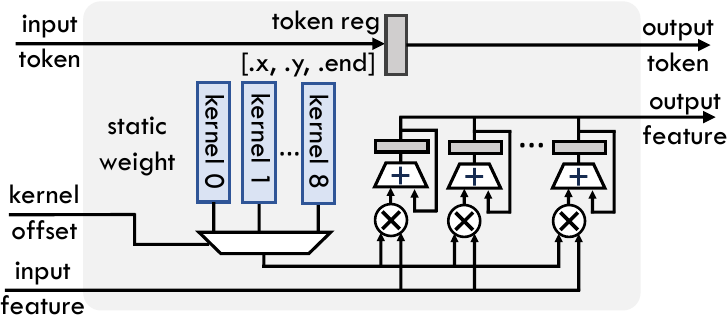}
    \centering
    \caption{Depthwise Convolution $3\times 3$ Computation Module.}
    \label{fig:conv3x3 compute}
\end{figure}
\figref{fig:conv3x3 compute} shows the diagram of the $3\times 3$ depthwise convolution module, which is widely used in many mobile networks~\cite{mobilenetv2, mnasnet, efficientnet}. 
For each non-zero output location marked by the current token register, the weighted sum of sparse neighboring features is controlled by the additional kernel offset stream. 
The kernel offset marks the relative spatial locations of the current input features in the computation window, which also serves as the index of the weights buffer. 
After performing the weighted sum for all non-zero kernel offsets, the final results will be streamed out as the output features. 
Other types of convolution, such as full convolution, only differ slightly in the PE organization, while having identical token interfaces and sparse weighted sum operations. 
Like the $1\times 1$ convolution module, the hardware template also incorporates the channel dimension parallel factor as a configurable variable.

\subsubsection{Sparse Line Buffer with Stride Equal to 1}
\begin{figure}[h]
    \includegraphics[page=4, width=0.9\linewidth]{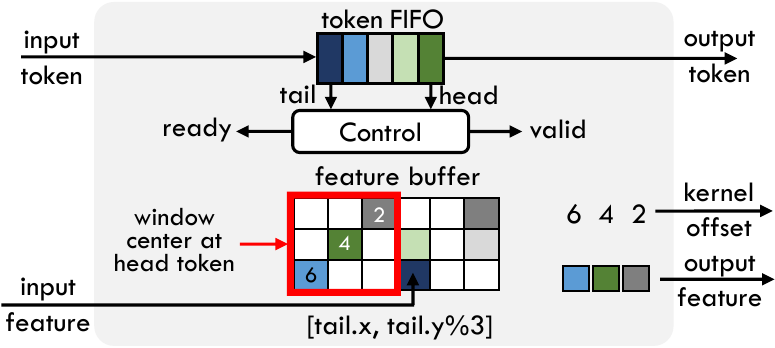}
    \centering
    \caption{Sparse Line Buffer Module (stride=1).}
    \label{fig:SLB s1}
\end{figure}
The Sparse Line Buffer (SLB) mainly comprises a $k$ rows data buffer ($k$ being the kernel size), a token FIFO, and a small bitmap to record current non-zero spatial locations in the buffer. 
When stride $s=1$, submanifold convolution enforces the input and output non-zero locations to be identical. As a result, a simple token FIFO is used to set up a queue for the output tokens.  
Due to the deterministic streaming order of the sparse sliding window algorithm (left-to-right and top-to-bottom as discussed in \eqnref{eq:dataflow}), we can simply use the token FIFO's head and tail tokens to control the read and write operations of SLB. 

Specifically, the FIFO's head token should represent the next spatial location for weighted sum computation, while the FIFO's tail token indicates the most recently received input location.
The offset between the head and tail token can be used to determine whether the buffer already contains all the data required for the next output (center at head token). If the current head token is $h$, the required input window would be $[h.x - u, h.x + u] \times [h.y - u, h.y + u]$, where $u=(k-1)/2$ is the kernel radius. Since our dataflow adheres to the top-to-bottom, left-to-right ordering, when the tail token $t$ points to a location with ravel order larger than $[h.x + u, h.y + u]$, it signifies that all the required data are already in the buffer. The head token is valid to be sent out, and the corresponding features are streamed for computation. 

In terms of the ready signal, new tokens and features are only allowed to be written into the SLB if the input still lies in the currently buffered rows/region. When the new input is beyond the buffered rows, SLB will first output the buffered tokens before accepting new inputs.
This approach is inherently free from deadlock as the arrival of a new token from subsequent rows indicates that tokens from preceding rows ($\leq t.y - u - 1$) are all valid for output. This is because the ravel order of $t$ should be larger than $[h.x + u, h.y + u]$ when $h.y \leq t.y - u - 1$. 
Those valid tokens will be popped out one by one, and buffers will be cleaned up to allow new input data to be buffered.
In summary, the control signals ready $r$ and valid $v$ can then be written as: 
\begin{equation}
\small
\begin{split}
    v&=((t.x-h.x\geq u) \wedge (t.y-h.y\geq u))\vee (t.y-h.y\geq u + 1)\\
    r&= (t.y-h.y\leq u)\\
    u&=(k-1)/2
\end{split}
\label{eq:elb}
\end{equation}
where $h,t$ stands for head and tail tokens. $u$ is the radius of the kernel. $k$ refers to kernel size.

As mentioned before, the kernel offset stream is used to denote each non-zero location within the $k\times k$ window. This information is obtained by querying a bitmap that marks whether a spatial location in the buffer is currently non-zero or not.
\figref{fig:SLB s1} shows an example diagram of an SLB for $3\times 3$ convolution with stride equal to 1. The snapshot shows the kernel window center at the head token, where the features on kernel offset $2, 4, 6$ are non-zeros. In this case, the features at $2,4,6$ will be outputted along with the kernel offsets for downstream weighted sum computation.

\subsubsection{Sparse Line Buffer with Stride Equal to 2}
\begin{figure}[h]
    \includegraphics[page=5, width=0.95\linewidth]{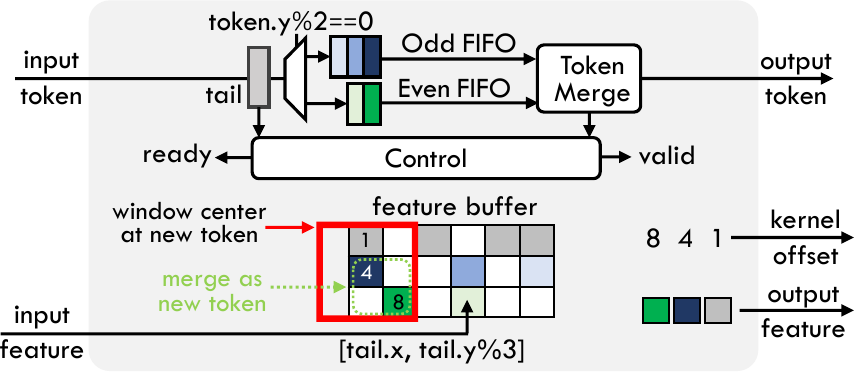}
    \centering
    \caption{Sparse Line Buffer Module (stride=2).}
    \label{fig:SLB s2}
\end{figure}
When using the stride $s=2$ in a convolution layer, the output feature map will be downsampled.
As illustrated in \figref{fig:submanifold}b, the sparse convolution will recalculate the non-zero coordinates (tokens). 
For each $2\times 2$ input stride grid, if it contains any non-zero pixels, it will generate a valid non-zero output. 
In this case, the Sparse Line Buffer module with stride equal to 2 (SLB s2) adopts two token FIFOs to buffer tokens from even and odd rows, respectively. A Token Merge unit is positioned at the head of the two FIFOs to calculate the new output tokens, which can be written as:
\begin{equation}
\label{eq:slb-s2}
\begin{split}
       &h\_{e2} = [h\_{e}.x/2,\ h\_{e}.y/2,\ h\_{e}.end]\\
       &h\_{o2} = [h\_{o}.x/2,\ h\_{o}.y/2,\ h\_{o}.end]\\
        \text{token}_\text{{out}} =& (ravel(h\_{e2}) > ravel(h\_{o2}))\ ?\ h\_{o2} : h\_{e2}
\end{split}
\end{equation}
where the $h\_{o}$ and $h\_{e}$ are the head tokens from odd and even token FIFOs, respectively. When a new output token is generated, the tokens in the corresponding $2\times2$ grid will all be popped out from FIFOs.

Similar to the case with stride being 1, the tail token $t$ can help determine whether the new output token already sees sufficient data in the buffer. The control logic is similar to \eqnref{eq:elb}, with the head token $h$ being the $\text{token}_\text{{out}}$ in \eqnref{eq:slb-s2}.
\figref{fig:SLB s2} demonstrated an example diagram of an SLB s2 module for $3\times 3$ convolution, where the dark blue and green positions within the $2\times 2$ stride grid signify a valid output. The token merge units will pop the dark blue and green tokens from both token FIFOs and stream out the new token. 
The SLB s2 module then outputs the corresponding features and kernel offsets ($1,4,8$) for downstream computation. 

In typical Deep Neural Network (DNN) model designs, the kernel stride values are commonly set to either 1 or 2.
If using an even larger stride size ($s>2$), the SLB can be designed similarly by splitting $s$ number of token FIFO channels. 


\subsubsection{Pooling and FC Module}
\begin{figure}[h]
    \includegraphics[page=6, width=0.95\linewidth]{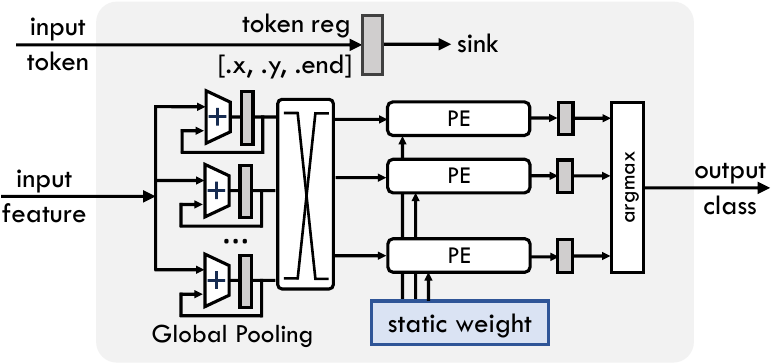}
    \centering
    \caption{Pooling and FC Module.}
    \label{fig:pooling and fc}
\end{figure}
For the object classification task, a model usually uses a global pooling layer followed by a fully connected layer to generate predictions. In the sparse dataflow of \aname, the global pooling layer will iterate through all the non-zero tokens until it encounters a token with the $.end=true$ flag. The features are aggregated into a single vector based on the pooling algorithm (maximum or average). Finally, the pooling results are fed into a linear classifier layer to compute the final prediction. \figref{fig:pooling and fc} shows an example diagram of a global average pooling layer combined with a fully connected classifier.

\subsubsection{Chaining Modules into a Block}
\begin{figure}[h]
    \includegraphics[page=7, width=0.95\linewidth]{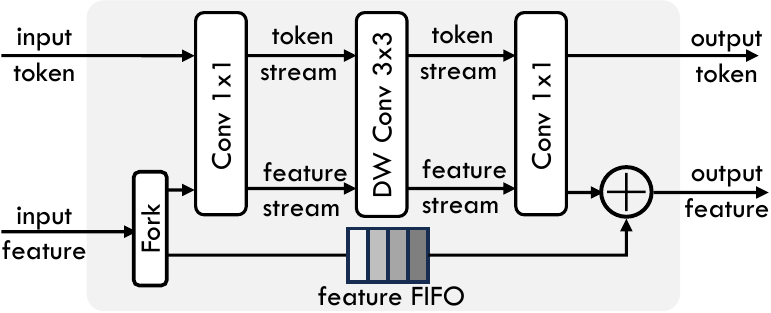}
    \centering
    \caption{Chaining a Residual Block.}
    \label{fig:conv block}
\end{figure}
Many DNN models are built upon convolution blocks with residual connections~\cite{resnet}. Thanks to the autonomous sparse iteration capabilities of all the modules in \aname, building convolution blocks by chaining the dataflow modules is straightforward. \figref{fig:conv block} shows an example of building an inverted residual block MBConv with stride being 1. MBConv is commonly employed in efficient models like MobileNetV2~\cite{mobilenetv2}, EfficientNet~\cite{efficientnet}, MnasNet~\cite{mnasnet}, etc.  Although \aname mainly uses MBConv blocks in the following experiments, similar approaches can be applied to other models like ResNet~\cite{resnet}, ShuffleNet~\cite{shufflenet}, etc. 
\figref{fig:conv block} illustrates a convolution block incorporating an identity shortcut, with all layer strides set to 1. Leveraging the principles of submanifold convolution, the input and output tokens remain identical across all the layers within this block. Thus, the input feature stream can simply be forked into two separate streams. One stream goes to the residual functional layers, while the other is used for the identity shortcut. 
The features in the shortcut branch are buffered in a FIFO and will be merged at the output of the last $1\times 1$ convolution layer.

\subsection{Customize Accelerator with Algorithm/Hardware Co-optimization}

The highly parameterized and composable sparse dataflow modules in \aname empower us to customize all-on-chip accelerators for different models or applications by spatially mapping distinct network components on FPGA. 
Moreover, this flexibility allows us to trade off between accuracy, resource consumption, and performance. However, it's essential to recognize that the design space can be vast for this problem. To address this challenge, \aname incorporates an algorithm/hardware co-optimization flow to automate this process.  

\subsubsection{Sparsity-awared Hardware Optimization}
In a dataflow accelerator, the throughput is usually bounded by the slowest module in the pipeline. Thus, finding a load-balanced hardware resource configuration for different modules is critical. 
To approach the optimal configuration, \aname adopts a static optimization strategy that hinges on application-wise sparsity statistics. More precisely, for each application/dataset, we collect the spatial and kernel sparsity ratios ($S_{s}$ and $S_{k}$) from all the samples in the dataset. The degree of spatial sparsity within the feature activation reflects the total number of iterations in each dataflow module, while the kernel sparsity indicates the averaged number of weighted sum operations occurring within convolution layers.
Using the sparsity statistics, we can estimate the average latency and resource utilization for each dataflow module. 
Take the depthwise $3\times3$ convolution as an example. The latency, the number of BRAM, and the DSP consumption can be estimated as:
\begin{equation}
\begin{split}
    lat &= (H \times W \times S_s)\times (9\times S_k) \times (C/PF) \\ 
    bram &= \lceil(B \times 9 \times C) / 16 \text{K} / PF\rceil \times PF\\
    dsp &= PF
\end{split}
\end{equation}
where $H,W$ is the input height and width in the current layer, $C$ is the number of channels, and $PF$ is the channel parallel factor. $B$ is the bitwidth of the quantized weight and we assume one BRAM can store 16 Kb data. 
The $(H \times W \times S_s)$ term reflects the averaged number of iterations on the spatial dimension of this layer and $(9\times S_k) \times (C/PF)$ estimates the averaged cycles needed for computing an output. The constant weight buffer is also partitioned to satisfy the on-chip bandwidth requirement determined by the parallelism factor $PF$. 

Using the similar approaches, we can formulate the latency and resource utilization for the remaining modules. Since all the layers are pipelined, we can estimate the theoretical throughput by determining the maximum latency across all layers.
Thus, the overall hardware optimization problem can be written as:
\begin{equation}
    \label{eq:opt}
    \begin{split}
        \min \quad  lat& \\
        \text{s.t.}\quad  lat_i &\leq lat,\quad i \text{ in all layers} \\
                         \sum_i r_{ij}& \leq R_j,\quad i\text{  in all layers, j in [BRAM, DSP] }
    \end{split}
\end{equation}
where $r_{ij}$ means the $j$ type resource consumed for layer $i$. $R$ is the total resource of each kind available. The current framework mainly concerns BRAM and DSP resources on FPGA, and the proposed optimization program will be solved by an open-sourced mixed integer programming solver. 

\begin{figure}[t]
    \includegraphics[width=\linewidth]{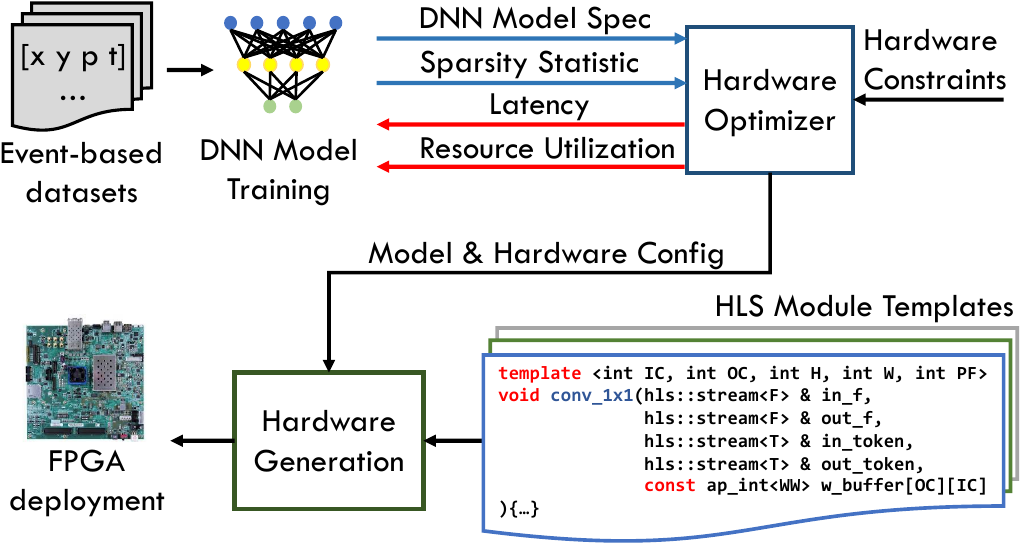}
    \centering
    \caption{Optimization flow of \aname.}
    \label{fig:eda}
\end{figure}

\subsubsection{Searching for Efficient Models}
\label{subsubsec:nas}
For an all-on-chip accelerator, the weights of all the layers should be placed on-chip using BRAM. 
In application scenarios with tight hardware resource constraints, deploying larger models like ResNet~\cite{resnet} might be infeasible. Many previous research works have explored integrating Neural Architecture Search into the deployment process to facilitate the model performances while satisfying the hardware constraints. 

Inspired by these concepts, we develop a simple two-step greedy search algorithm to identify efficient models for deployment. 
Initially, our software randomly samples a pool of model architectures built upon the MBConv convolution block. 
The search space encompasses variations in: (1) the number of blocks; (2) the stride of the block; (3) the number of channels in each layer. 
The models are sampled within a coarse-defined model size range according to the available on-chip buffer size. Also, the total down-sampling ratio of each model remains fixed.
These sampled models, along with the sparsity statistic, are fed into the hardware optimizer, which searches for the optimal hardware mapping of each model and estimates its performance. 
Running the hardware optimizer for each model typically takes only several seconds/minutes to complete. This means that for each dataset, we can effortlessly sample hundreds of models.

Once all the hardware performance profiles for all the models are collected, the software selects the top-k highest throughput models for training and finally chooses the model with the best accuracy. 
This approach allows us to identify models that fit the application constraints while balancing between accuracy and hardware performance.
Note that the current simple model search algorithm does not leverage any sophisticated reinforcement learning or other meta-learning algorithms. Nevertheless, the possibility of integrating other Neural Architecture Search algorithms presents an exciting direction for future exploration.
\figref{fig:eda} provides a summary of the co-optimization process in \aname. 
\section{Evaluations} \label{sec:evaluations}

\begin{figure*}[t]
    \includegraphics[width=\linewidth]{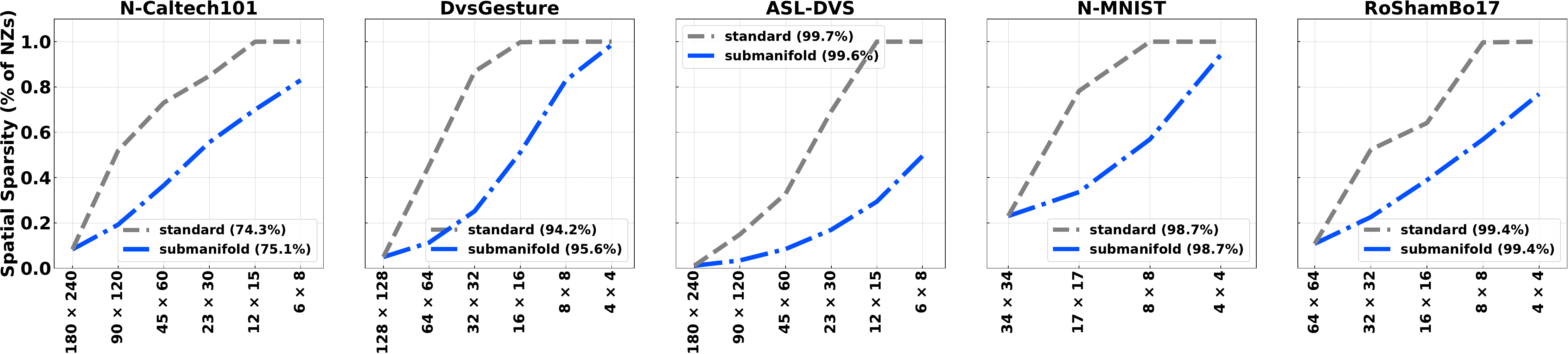}
    \centering
    \caption{
    Compare the spatial sparsity between standard and submanifold convolution on different event-based vision datasets. The x-axis of each subplot is the spatial resolution of the feature activation. Accuracy results of two different models in floating point precision are also written in the legends. 
    }
    \label{fig:sparsity statistic}
\end{figure*}


\subsection{Experiment Setup}
We evaluated \aname on five different event-based gestures or object recognition datasets: DvsGesture~\cite{dvsgesture}, RoShamBo17~\cite{roshambo}, ASL-DVS~\cite{asl},  N-MNIST and N-Caltech101~\cite{ncal}. DvsGesture contains 10 classes of human body gestures captured by DVS128 camera and RoShamBo17 records hand gestures playing rock-scissors-paper games. ASL-DVS is a large-scale event-based American sign language letters recognition dataset recorded by DAVIS240C event camera. 
N-Caltech101 and N-MNIST datasets slightly differ from previous datasets that were directly shot in real scenes. Instead, both are recaptured from static image-based Caltech101~\cite{cal101} and MNIST~\cite{mnist} datasets by using an actuated pan-tilt event-camera platform. All the datasets are preprocessed with a similar method that first clips event recordings with a fixed time interval and then generates histogram representation with two channels (positive and negative events). Except for the RoShamBo17 dataset, we directly use their preprocessed histogram format. 

For hardware experiments, we implement \aname on an embedded FPGA board ZCU102 with a Zynq UltraScale+ MPSoC device (XCZU9EG) using Vitis HLS and Vivado Design Flow. End-to-end hardware performances are measured with on-board testing after synthesis, place, and route. Models are trained using PyTorch and MinkowskiEngine~\cite{mink} sparse DNN library. Weights and activations are quantized into 8 bits by using HAWQ-V3~\cite{hawqv3} framework in our training pipeline.    
The hardware optimization program in \eqnref{eq:opt} is solved by a mixed-integer geometric programming solver developed in AGNA~\cite{agna} with SCIP~\cite{scipopt} and GPkit~\cite{gpkit} optimization frameworks.

\subsection{Performance of Submanifold Convolution}

\figref{fig:sparsity statistic} shows the spatial statistics from different datasets. The sparsity ratios are measured on different feature resolutions throughout the network. For example, the N-MNIST dataset has the input resolution being $34\times 34$. Through the downsample stages of the network, the resolution will change to $17\times 17$, $8\times 8$, and finally $4\times 4$. 
We can see that submanifold sparse convolution can significantly improve sparsity compared to standard convolution. Even though all the input data are very sparse (1.1\% to 23.1\%), the dilation effect (shown in \figref{fig:submanifold}) of standard convolution can easily lead to very dense features throughout the network. 
In ASL-DVS dataset, using submanifold sparse convolution can lead to as much as $3.4\times$ sparser features than standard convolution. 

To compare the accuracy performances of two different types of convolution, we trained MobileNetV2 models with both convolution methods on N-Caltech101, DvsGesture, and ASL-DVS. 
Since N-MNIST and RoShamBo17 have smaller input resolutions, we use a customized network architecture instead. 
The accuracy results (written in the legends of \figref{fig:sparsity statistic}) show that for most datasets, submanifold sparse convolution achieves similar or higher accuracy compared with standard convolution. 
In summary, submanifold sparse convolution showcases its ability to effectively leverage the intrinsic spatial sparsity present in event-based datasets while having similar accuracy performances compared with standard convolution, making it a highly promising candidate for hardware acceleration.



\subsection{Compare \aname with Dense Dataflow}
\begin{figure*}[t]
    \includegraphics[width=\linewidth]{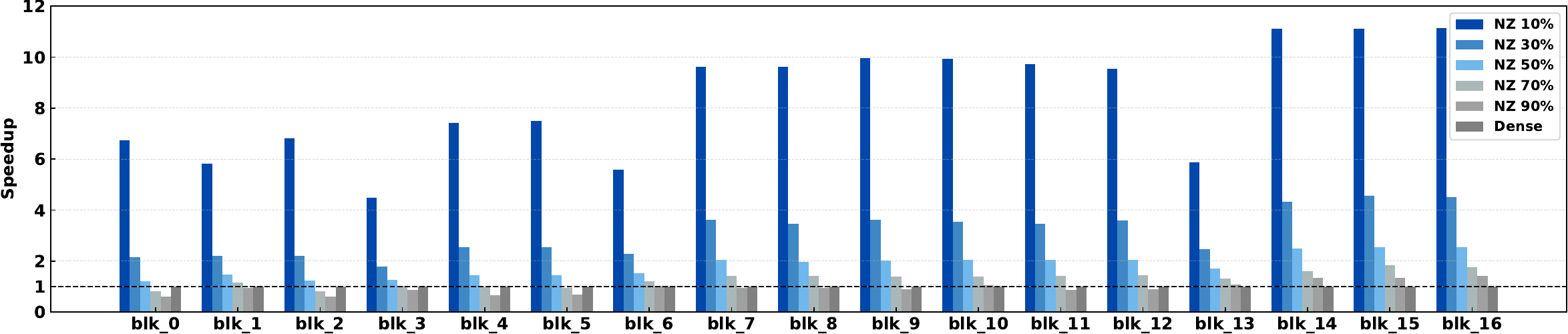}
    \centering
    \caption{Speedup of ESDA sparse dataflow architecture over sliding-window-based dense dataflow architecture with different input sparsity ratios (randomly generated). The sparse and dense designs use identical hardware configurations like Parallel Factor, Bitwidth, etc.  
    blk\_x refers to different convolution blocks in MobileNetV2.}
    \label{fig:benchmark}
\end{figure*}

To demonstrate the effectiveness of the sparse dataflow modules, we individually synthesized all the MBConv convolution blocks of the MobileNetV2 network and evaluated their performance across various input sparsity ratios. 
The hardware configuration of each block aligns with the overall optimization result of the entire MobileNetV2 architecture.
To evaluate the overhead for supporting dynamic sparse control logic, we implement a dense baseline with a similar sliding-window-based dataflow that uses a standard line buffer design. The dense baseline eliminates all token stream interfaces and dynamic logic components, maintaining identical parallel factors, bitwidths, and other design parameters in alignment with the sparse implementations. 

\figref{fig:benchmark} presents the speedup ratio of \aname modules over the dense baseline using C/RTL co-simulation latency. The inputs consist of randomly generated data with varying spatial sparsity ratios, ranging from 10\% to 90\%. The results demonstrate that dynamic sparse dataflow architecture can effectively leverage sparsity in general. With an input non-zero ratio of 10\%, the sparse designs achieve significant speedups of $4.5-11\times$ compared to the dense baselines. Moreover, the speedup shows a nearly linear increase as sparsity grows.

However, it is worth noting that the overhead of supporting dynamic sparse computation does exist. Specifically, when the input is nearly dense (greater than 70\% NZ), certain blocks like blk\_0 to blk\_5 exhibit slower performance than the dense baseline. These blocks typically have large spatial resolutions and smaller channels, making them more susceptible to overhead.
Fortunately, it turns out that these larger resolution layers in the early stages of a network typically have higher sparsity (usually below 40\% NZ) as shown in \figref{fig:sparsity statistic}. As a result, despite potential slowdowns with very dense input, these modules can still attain notable speedup with highly sparse inputs.

\subsection{Compare \aname with Embedded GPU}
\begin{figure}[t]
    \includegraphics[width=\linewidth]{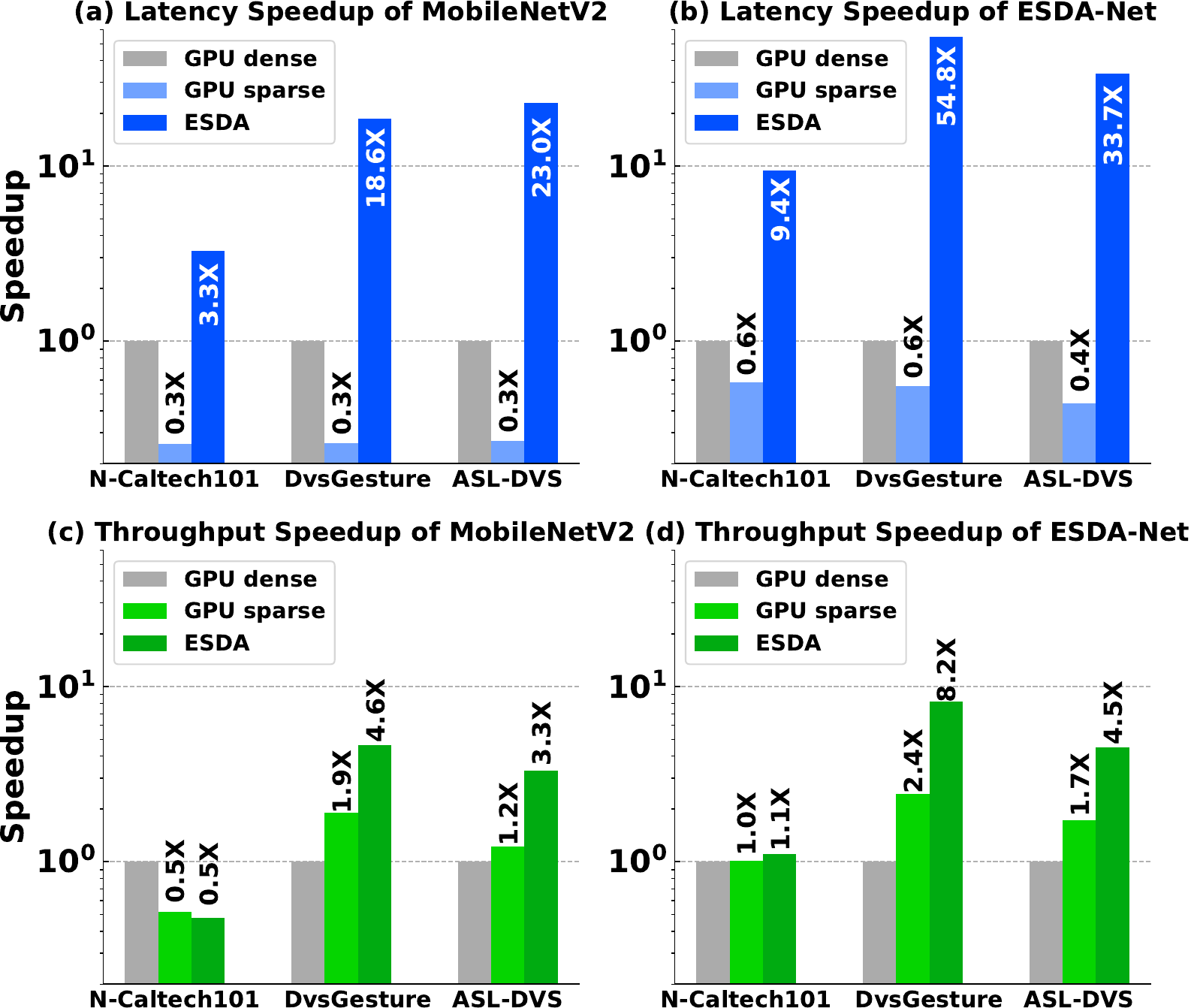}
    \centering
    \caption{Compare \aname with an embedded GPU platform (NVIDIA Jetson Xavier NX). For GPU performances, we tested both dense DNN implementation using PyTorch (baseline) and submanifold sparse DNN implementation using MinkowskiEngine. And we compared both MobileNetV2 and the customized models searched by \aname. }
    \label{fig:gpu}
\end{figure}

\figref{fig:gpu} shows the comparison between \aname and an embedded GPU platform (NVIDIA Jetson Xavier NX) on three event-based datasets: N-Caltech101, DvsGesture, and ASL-DVS. 
For GPU implementation, we tested the performance of (1) dense DNN implementation using PyTorch and (2) submanifold sparse DNN implementation using MinkowskiEngine~\cite{mink} library. 
In addition, we evaluated the MobileNetV2 (with width-multiplier being 0.5) as well as the customized ESDA-Net (the models searched by using \aname optimization flow for each dataset). The latency and throughput performances of GPU use batch size 1 and 128 respectively. 

In terms of latency performance, \aname exhibits a substantial speedup over both dense and sparse GPU implementations for all the model-dataset combinations. In comparison to the dense GPU baseline, \aname delivers $3.3-23.0\times$ speedup on MobileNetV2 and $9.4-54.8\times$ speedup on customized models. 
An interesting observation is that the latency performance of sparse GPU implementation lags behind the dense GPU baseline. This discrepancy primarily comes from the way that GPUs handle sparsity in submanifold convolutions. It stores the sparse input coordinates in a hash table and breaks the original computation into a fine-grained sparse gather-scatter flow for each spatial kernel (marked as k0-k8 in \figref{fig:submanifold}). In scenarios with relatively small batch sizes, this additional overhead can have an adverse impact on performance.

As for throughput performances, \aname surpasses the dense GPU baseline in most cases, with the exception being MobileNetV2 on N-Caltech101. The smaller speedup ratio is attributed to the fact that current \aname is mainly optimized for batch-1 low-latency inference. 
This design choice is driven by its primary focus on near-sensor or edge systems, where latency performance is vital. Nevertheless, our customized models searched by the co-optimization flow also demonstrate better speedup performances than MobileNetV2, which highlights the advantages and effectiveness of the framework. Last but not least, ESDA demonstrates an average of $5.8\times$ and $3.3\times$ improvement of energy efficiency (measured in Joules per inference) when compared with GPU dense and sparse implementations, respectively, across the benchmarks. This underscores the low-power capability of our sparse architecture for event-based vision DNN inference.

\begin{table*}[]
\caption{System Performance}
\centering
\footnotesize
\begingroup
\renewcommand{\arraystretch}{1.25}
\begin{tabular}{cccccccccccccc}
\hline
\multicolumn{1}{l}{\multirow{3}{*}{}} &
  \multirow{3}{*}{\textbf{Dataset}} &
  \multirow{3}{*}{\textbf{Resolution}} &
  \multirow{3}{*}{\textbf{Model}$^\text{a}$} &
  \multirow{3}{*}{\begin{tabular}[c]{@{}c@{}}\textbf{Bit-}\\ \textbf{width}\end{tabular}} &
  \multirow{3}{*}{\begin{tabular}[c]{@{}c@{}}\textbf{Acc}\\ \textbf{(\%)}\end{tabular}} &
  \multirow{3}{*}{\begin{tabular}[c]{@{}c@{}}\textbf{Latency}\\ \textbf{(ms)}\end{tabular}} &
  \multirow{3}{*}{\begin{tabular}[c]{@{}c@{}}\textbf{Throughput}\\ \textbf{(fps)}\end{tabular}} &
  \multirow{3}{*}{\begin{tabular}[c]{@{}c@{}}\textbf{Power}\\ \textbf{(W)}\end{tabular}} &
  \textbf{Energy} &
  \multicolumn{4}{c}{\multirow{2}{*}{\begin{tabular}[c]{@{}c@{}}\textbf{Implementation} \\ \textbf{Details}$^{\text{b}}$\end{tabular}}} \\
\multicolumn{1}{l}{} &
   &
   &
   &
   &
   &
   &
   &
   &
  \textbf{Efficiency} &
  \multicolumn{4}{c}{} \\ \cline{11-14} 
\multicolumn{1}{l}{} &
   &
   &
   &
   &
   &
   &
   &
   &
  \textbf{(mJ/inf)} &
  \textbf{DSP} &
  \textbf{BRAM} &
  \textbf{FF} &
  \textbf{LUT} \\ \hline
\multirow{8}{*}{\begin{tabular}[c]{@{}c@{}}\textbf{ESDA}\\ (Ours)\end{tabular}} &
  \multirow{2}{*}{N-Caltech101} &
  \multirow{2}{*}{$180\times240$} &
  ESDA-Net &
  8 &
  72.4 &
  3.09 &
  323 &
  1.81 &
  5.61 &
  1792 &
  1278 &
  115K &
  154K \\
 &
   &
   &
  MobileNetV2 &
  8 &
  71.6 &
  7.12 &
  140 &
  2.10 &
  14.96 &
  1992 &
  1600 &
  198K &
  207K \\ \cline{2-14} 
 &
  \multirow{2}{*}{DvsGesture} &
  \multirow{2}{*}{$128\times128$} &
  ESDA-Net &
  8 &
  92.5 &
  0.66 &
  1526 &
  1.58 &
  1.03 &
  1532 &
  848 &
  97K &
  128K \\
 &
   &
   &
  MobileNetV2 &
  8 &
  93.9 &
  1.19 &
  839 &
  1.73 &
  2.06 &
  1636 &
  1134 &
  104K &
  140K \\ \cline{2-14} 
 &
  \multirow{2}{*}{ASL-DVS} &
  \multirow{2}{*}{$180\times240$} &
  ESDA-Net &
  8 &
  99.5 &
  0.71 &
  1406 &
  1.60 &
  1.14 &
  1494 &
  917 &
  97K &
  131K \\
 &
   &
   &
  MobileNetV2 &
  8 &
  99.3 &
  1.08 &
  927 &
  1.75 &
  1.88 &
  1416 &
  1069 &
  108K &
  144K \\ \cline{2-14} 
 &
  N-MNIST &
  $34\times34$ &
  ESDA-Net &
  8 &
  98.9 &
  0.15 &
  6657 &
  1.55 &
  0.23 &
  1525 &
  978 &
  93K &
  121K \\ \cline{2-14} 
 &
  RoShamBo17 &
  $64\times64$ &
  ESDA-Net &
  8 &
  99.6 &
  0.98 &
  1016 &
  1.40 &
  1.38 &
  1282 &
  765 &
  72K &
  95K \\ \hline
NullHop &
  RoShamBo17 &
  $64\times64$ &
  RoshamboNet &
  16 &
  99.3 &
  10 &
  160 &
  0.27 &
  1.69 &
  657 &
  802 &
  139K &
  266K \\ \hline
PPF &
  - &
  $60\times40$ &
  PFF-BNN &
  1 &
  87 &
  7.71 &
  - &
  - &
  - &
  12 &
  223 &
  16K &
  42K \\ \hline
Asynet &
  N-Caltech101 &
  $180\times240$ &
  VGG &
  FP32 &
  74.5 &
  80.4 &
  - &
  - &
  - &
  \multicolumn{4}{c}{CPU} \\ \hline
TrueNorth &
  DvsGesture &
  $64\times64$ &
  SNN &
  Tenary &
  94.6 &
  105 &
  - &
  0.18 &
  18.7 &
  \multicolumn{4}{c}{Samsung 28 nm LPP CMOS process} \\ \hline
Loihi &
  DvsGesture &
  $32\times32$ &
  SNN &
  9 &
  90.5 &
  11.43 &
  - &
  - &
  - &
  \multicolumn{4}{c}{Intel 14 nm process} \\ \hline
\multicolumn{14}{l}{$^{\text{a}}$ ESDA-Net refers to the customized network for each dataset. $^{\text{b}}$ If written in 4 columns, the designs are implemented on FPGA.}
\end{tabular}
\endgroup
\label{tab:system_perf}
\end{table*}

\subsection{Compare \aname with Other Related Works}

\tabref{tab:system_perf} provides the detailed hardware implementation results of \aname and other related works on different datasets. The hardware performances of \aname were measured on a ZCU102 FPGA board with the programmable logic (PL) running at \SI{187}{\MHz}. The power was measured during runtime using the built-in power monitor controlled by the PMBus. 
For N-Caltech101, DvsGesture, and ASL-DVS datasets, we implement two models: MobileNetV2 (with the width-multiplier being 0.5) and the customized models. Note that the customized models are all named EDSA-Net in \tabref{tab:system_perf} for brevity but the model for each dataset is different. 

In general, \aname demonstrates superior performance in terms of overall system efficiency and effectiveness. Across all the datasets, \aname achieves ultra-low latency (\SI{0.15}{\ms} to \SI{7.12}{\ms}), significantly boosting the real-time performance of event-based DNN inference. 
As mentioned before, the current implementation of \aname only supports batch size 1 inference, and the throughput is calculated using 1/latency. Nevertheless, \aname still achieves over 1000 fps throughput for most datasets except N-Caltech101. 
The performance of N-Caltech101 dataset is slightly lower than the others because it has a larger spatial resolution and denser input. 
On the other hand, \aname also showcases notable low power and high energy efficiency features. The power consumption of the FPGA fabric ranges from \SI{1.4}{\W} to \SI{2.1}{\W} across different datasets, with the average energy per inference spanning from \SI{0.23}{\mJ} to \SI{14.96}{\mJ}. In summary, \aname provides a promising solution for low-power DNN acceleration on edge event-based vision systems.

\aname also demonstrates better system performances compared to other prior FPGA accelerators, NullHop~\cite{nullhop, nullhop_fpga} and PPF~\cite{ppfbnn}. Both are FPGA DNN accelerators that have been applied to event-based vision tasks. On the RoShamBo17 dataset, \aname achieves $10.2\times$ speedup in latency and $6.4\times$ speedup in throughput while also having better accuracy performance. 
In terms of power efficiency, although NullHop exhibits lower power consumption, which may be attributed to its lower clock frequency (\SI{60}{\MHz}), \aname outperforms it in terms of energy efficiency (\SI{1.38}{\mJ/inf} vs \SI{1.69}{\mJ/inf}). 
PPF is a binary neural network FPGA accelerator for event-based pedestrian detection. Since PPF did not release their datasets, we can not compare with them directly. With similar input resolution on the RoShamBo17 dataset, \aname has significant latency speedup over PPF (\SI{0.98}{\ms} vs \SI{7.71}{\ms}).

Asynet~\cite{asynet} is an asynchronous event-based submanifold sparse convolution network implemented on CPU. Despite the attractive asynchronous feature of Asynet, the overhead of asynchronous computation can be high, which involves updating the kernel maps and feature vectors incrementally for each event. 
On the N-Caltech101 dataset, \aname achieves $26\times$ speedup compared with Asynet with only around 2\% lower accuracy. When implementing Asynet on customized hardware, one can expect a significant amount of random memory access for fetching and updating non-zero features. Otherwise, the accelerator shall buffer all the intermediate features on-chip, which can easily overwhelm the available resources. 

We also compare \aname with neuromorphic hardware solutions. TrueNorth~\cite{truenorth_chip} and Loihi~\cite{loihi} are well-known SNN chips, and they both have been used to accelerate event-based gesture recognition on the DvsGesture dataset~\cite{dvsgesture, loihi_gesture}. In terms of latency, \aname achieves $160\times$ and $17.4\times$ speedup over TrueNorth and Loihi, respectively. At the same time, \aname also exhibits $18\times$ better energy efficiency compared with TrueNorth. In essence, \aname attains rapid and energy-efficient DNN acceleration for event-based vision without relying on traditional neuromorphic techniques, which paves the way to future efficient event-based systems. 



\section{conclusion} 
\label{sec:conclusion}

In this work, we have presented \aname, a novel composable dynamic sparse dataflow architecture for efficient event-based vision DNN inference. 
\aname confronts the challenge of low-latency and low-power processing of event-based data by focusing on its inherent sparsity. This is made possible by synergizing algorithmic advancements through submanifold sparse convolution with novel architectural solutions realized through the sparse dataflow design.
The flexibility and programmability of the \aname are also supercharged by the highly modular dataflow design as well as the sparsity-aware design optimization flow for FPGA platforms. 
Extensive experiments show the \aname achieves up to $54.8\times$ speedup than the embedded GPU as well as $10.2\times$ speedup compared to NullHop. In addition, \aname also shows $160\times$ speedup and $17.4\times$ better energy efficiency compared to SNN hardware platform TrueNorth.   
The ability to effectively exploit sparsity in \aname opens up new possibilities for real-time, energy-efficient DNN inference for event-based vision applications.

\begin{acks}
This work was supported in part by the Research Grants Council (RGC) of Hong Kong under the Research Impact Fund project R7003-21 and the Theme-based Research Scheme (TRS) Project T45-701-22-R. This work was supported by AI Chip Center for Emerging Smart Systems (ACCESS), sponsored by InnoHK funding, Hong Kong SAR. 
\end{acks}

\bibliographystyle{ACM-Reference-Format}
\bibliography{ref}










\end{document}